\documentclass[a4paper,11pt]{article}
\usepackage[utf8]{inputenc}
\usepackage[margin=1in]{geometry}  

\usepackage{graphicx}
\usepackage{gensymb}
\usepackage{tabularx}
\usepackage{aas_macros}
\usepackage{adjustbox}
\usepackage{subcaption}
\usepackage{fancyhdr}
\usepackage{titling}  
\usepackage{ragged2e}
\usepackage[normalem]{ulem} 
\usepackage[colorlinks =TRUE,
            urlcolor  = blue,
            citecolor = blue,
            anchorcolor = blue]{hyperref}    
\usepackage{natbib}
\usepackage{listings}
\usepackage{lipsum} 

\makeatletter
\renewcommand{\maketitle}{
  \begin{flushleft}
    {\textit{\small WDS'24 Proceedings of Contributed Papers — Physics, 178–184, 2024. \\
  ISBN 978-80-7378-520-8 © MATFYZPRESS.}}
    \vspace{1em} 
    
    {\LARGE\bfseries \@title \par}
    \vspace{2em}
    {\large \@author \par}
    \vspace{0.5em}
    {\small \@date \par}
  \end{flushleft}
}
\makeatother

\title{\textbf{X-ray polarization of accreting black holes: \\ Cyg X--1 and Swift J1727.8--1613}}
\author{M. Brigitte and J. Svoboda \\ \small{Astronomical Institute of the Czech Academy of Sciences.}}
\date{}  

\begin{document}

\maketitle

\begin{flushleft}
    {\large \textbf{Abstract.}  The Imaging X-ray Polarimetry Explorer is an X-ray observatory measuring the X-ray polarization in the 2--8 keV energy range. Highly sensitive to the system's geometry, X-ray polarization is a unique method to probe the structure of X-ray binaries. The Imaging X-ray Polarimetry Explorer observed the High-Mass X-ray Binary Cygnus X--1 and the Low-Mass X-ray Binary Swift J1727.8--1613 in different accretion states: in the hard state and in the soft state. The X-ray polarimetry analysis of both sources shows a linear polarization degree increasing with energy, with higher values in the hard state than in the soft state. However, the linear polarization angle stays similar in both states and is aligned with the radio jet within 5\degree. Furthermore, the Low-Mass X-ray Binary Swift J1727.8--1613 has a lower optical intrinsic polarization and a lower X-ray polarization degree for a softer spectrum. The similarities observed in this analysis between the X-ray polarization results of different types of X-ray Binaries show that the innermost accretion processes are independent of the companion star's type.  }\\[1ex]
\end{flushleft}

\section{Introduction}
X-ray Binaries (XRBs) contain a star and a compact object (a neutron star or a black hole) orbiting the system's centre of mass. Such systems are called High-Mass X-ray Binaries (HMXBs) or Low-Mass X-ray Binaries (LMXBs) depending on the companion star's mass. When the compact object accretes matter from the star, an accretion disk forms around it, and the matter gets heated to millions of degrees, releasing an intense X-ray emission. XRBs are variable in X-rays and transition from a so-called "hard state" to a "soft state" and vice versa. In the soft state, the emission is dominated by the thermal multi-blackbody emission, well represented by the standard solution of the accretion disk by \cite{1973A&A....24..337S}. In the hard state, the emission is dominated by the up-scattered emission in a hot Comptonized plasma, also called the "corona" \cite[][]{1975ApJ...195L.101T, 1976ApJ...204..187S, 1979Natur.279..506S}. In this state, the reflection of the hot photons from the corona on the accretion disk generates the iron ($\mathrm{K\alpha}$) fluorescence line at 6.4 keV and a Compton hump around 20 keV \cite[][and references therein]{2010ApJ...718..695G}. The shape and position of the corona is still debated, and the X-ray spectra can be fitted with different models \cite[][]{1976ApJ...204..187S, 1996ApJ...470..249P, 2021SSRv..217...65B}. 

Cygnus X--1 (Cyg X--1 or HDE 226868) is one of the most persistent HMXBs in our Galaxy. The system hosts a $\mathrm{ 40.6 ^{+ 7.7}_{- 7.1} \ M_{\odot}}$ blue supergiant and a $\mathrm{21.2 \pm 2.2 \ M_{\odot}}$ intermediate-mass black hole at $\mathrm{2.22^{+ 0.18}_{-0.17}}$ kiloparsecs (kpc) \cite[][]{2021Sci...371.1046M}. The massive star emits strong stellar winds, captured and accreted by the black hole. The quasi-circular orbit has an inclination of 27.1\degree ± 0.8\degree \cite[][]{2011ApJ...742...84O, 2021Sci...371.1046M} and an orbital period of 5.599829 ± 0.000016 days \cite[][]{1999MNRAS.309.1063B, 2011ApJ...742...84O}. Cyg X--1 rarely reaches the complete soft state but remains in a soft-intermediate state where the coronal emission is still present. Swift J1727.8--1613 is a LMXB that went into a bright outburst and was detected on August 24, 2023 \cite[][]{2023GCN.34540....1K}. Since then, multiwavelength campaigns followed the source over multiple X-ray spectral states, including the Imaging X-ray Polarimetry Explorer \cite[see][]{2023ApJ...958L..16V, 2024A&A...686L..12P, 2024ApJ...966L..35S, 2024ApJ...968...76I}. Optical observations suggest  the presence of a black hole and an early K-type companion star, with an orbital period of approximately 7.6 hours at a distance of 2.7 ± 0.3 kpc \cite[][]{2024A&A...682L...1M}. High-resolution radio observations with the Very Long Baseline Array and the Long Baseline Array show the presence of a continuous radio jet along the black hole rotational axis \cite[][]{2024ApJ...971L...9W}. 

\indent The first X-ray polarimetry results from the Imaging X-ray Polarimetry Explorer \cite[IXPE, 2--8 keV,][]{2022JATIS...8b6002W, 2022JATIS...8b4003R} unravelled part of the mystery. X-ray polarization is characterized by the Polarization Degree (PD) and the Polarization Angle (PA). The PD quantifies the fraction of the X-ray light that has been polarized, and the PA specifies the orientation of the polarization with respect to a direction of reference. The first IXPE observations of Cyg X--1 showed a relatively high PD of 4.01\% ± 0.20\% in the hard state \cite[][]{2022Sci...378..650K} and a lower PD of 1.99\% ± 0.13\% (68\% confidence) in the soft state \cite[][]{2024ApJ...969L..30S}. On the other hand, the PA does not vary with energy and remains constant at -20.7\degree ± 1.4\degree \ in the hard state \cite[][]{2022Sci...378..650K} and -25.7\degree ± 1.8\degree \ East of North in the soft state \cite[68\% confidence,][]{2024ApJ...969L..30S}. The PA is aligned with the radio emission \cite[][]{2022Sci...378..650K, 2024MNRAS.52710837J, 2024ApJ...969L..30S}, believed to come from a relativistic jet located along the rotational axis of the black hole \cite[][]{2006MNRAS.369..603F}. Moreover, X-ray polarimetry is highly sensitive to the system's geometry. The first IXPE results indicate a corona extended perpendicular to the jet axis \cite[][]{2022Sci...378..650K}.
\cite{2024ApJ...969L..30S} observed an increase in the PD with the energy scaled by the disk temperature in both states. This interesting result shows that the PD energy spectra have a similar trend in both states despite completely different polarization origins and timing characteristics.\\
In comparison to the X-ray polarization produced in the innermost parts of the accretion disk, the optical polarization probes the radiation from the companion star at larger scales \cite[][]{2023A&A...678A..58K}. Therefore, comparing the two different types of polarization can give some insight into the influence of the companion star on the polarization properties of the source in different energy ranges.\\
\indent In this analysis, we first show the spectral differences between the different states of the HMXB Cyg X--1, using some data already reduced from \cite{2024ApJ...969L..30S} and additional complementary data. Then, we compare the polarization properties of Cyg X--1 with the ones from the newly discovered transient Swift J1727.8--1613. Comparing the polarization properties of the HMXB with the ones from a LMXB in different energy bands could give some insight into the different emission processes at play with different donors.

\section{Observations and data reduction}
IXPE observed Cyg X--1 in the hard state in May 2022 with a total exposure time of 242 ks. The same source was observed a year later in the soft-intermediate state in May-June 2023 with a total exposure time of 140 ks (see Table \ref{tab: obs log}). We downloaded the processed {\it level-2} data from the HEASARC archive for the three IXPE detectors. The source region is filtered with the \textit{xpselect} tool from the IXPEOBSSIM software package version 30.6.4 \cite[][]{2022SoftX..1901194B}. The source is defined as a circle with a radius of $80 ^{\prime\prime}$. The background region is defined as an annulus centered on the source, with an inner radius of $150 ^{\prime\prime}$ and an outer radius of $310 ^{\prime\prime}$. \\
Simultaneous X-ray observations were performed with the Neutron Star Interior Composition Explorer Mission \cite[NICER, 0.2--12 keV, see][]{2012SPIE.8443E..13G}. 
NICER monitored Cyg X--1 seven times in the hard state and four times in the soft state. The respective exposure times are approximately 90 ks and 35 ks (see Table \ref{tab: obs log}). The data were processed and filtered using the NICERDAS pipeline version 12. The data were screened to remove noisy detectors and intervals of high background, and the SCORPEON model was used for the background estimation \cite[][]{2022AJ....163..130R, 2024ApJ...969L..30S}. While the {\it nicerl2} task generated the {\it level-2} files, the {\it nicerl3} task generated the spectra and light-curves products in the 0.5--11 keV energy range because of the high noise below 0.5 keV. In the hard state, we combined the 5 observations from NICER using the task {\it niobsmerge} from NICERDAS. \\
Complementary to NICER, Cyg X--1 was observed with the Nuclear Spectroscopic Telescope Array \cite[NuSTAR, 3--79 keV,][]{2013ApJ...770..103H}.
NuSTAR monitored Cyg X--1 three times in the hard state and three times in the soft state (see Table \ref{tab: obs log}). The respective total exposure times are approximately 42 ks and 33 ks. The data were reduced using the {\it nupipeline} and {\it nuproducts} modules from NuSTARDAS. In the hard state, we combined the 3 observations from NuSTAR using the task {\it ftgrouppha} \cite[][]{2016A&A...587A.151K} from NuSTARDAS. In the soft state, the observations were divided into 5 epochs to avoid saturation \cite[see][and Table \ref{tab: obs log}]{2024ApJ...969L..30S}. \cite{2020arXiv200500569M} reported a thermal blanketing on the rear of the Focal Plane Modules (FPM)A, which generates differences in the flux with FPMB. Therefore, a multi-layer insulation (MLI) correction was applied between FPMA and FPMB. For FPMA, the offset is fixed at 0.964657 and the nuMLI covering factor at 0.999654. Both parameters are fixed at 1 for FPMB. We added a normalization constant to account for calibration discrepancies between the different instruments. The constant was fixed at 1 for NICER data and free for NuSTAR FPMA. The constant for NuSTAR FPMB is fixed to the one of NuSTAR FPMA. \\


\begin{table}[h!]     
    \caption{Observation log of IXPE, NICER and NuSTAR in the hard and soft state of Cyg X--1. The orbital phases are calculated from the ephemeris from \cite{2003ApJ...583..424G}.}
    \centering 
    \label{tab: obs log}
    \begin{tabular}{ c c c c c c}  
    \hline\hline \\ [-2.5ex]                  
    Date 	&	Telescope &  ObsID & Orbital phase $\mathrm{\phi}$ & Exposure time & Spectral state    \\  [0.08cm]  
		[UTC]	&	& 		&	& [ks]	 &  \\ [0.08cm]
    \hline
		2022 May 15	&  IXPE	&	01002901 &	0.77	&	327.0	& hard\\	[0.04cm]
					& NICER & 	5100320101 & 	& 	8.1  &   hard\\	[0.1cm]
		2022 May 16	& NICER &  5100320102   &  0.94  & 3.7 	& hard\\	[0.04cm]
					& 		 &  5100320103  &   & 23.6	&  hard\\	[0.1cm]
		2022 May 18	& NICER	 &  5100320104  & 0.30  & 26.7	& hard\\	[0.04cm]
					& 		 &  5100320105  &   & 14.4	&  hard\\	[0.04cm]
					& NuSTAR &  30702017002 &   & 16.2 &  hard\\	[0.1cm]
		2022 May 19 & NuSTAR &  30702017004 & 0.48  & 13.9 & hard\\	[0.1cm]
		2022 May 20  & NICER	&  5100320106   & 0.66  & 11.1	& hard\\	[0.04cm]
					& NuSTAR &  30702017006 &   & 12.4 &  hard\\	[0.1cm]
		2022 May 21  & NICER	&  5100320107   & 0.84  & 3.0 	& hard\\	[0.04cm]
		2022 June 18	&  IXPE	&	01250101 &	0.84	&	86.1	& hard\\	[0.4cm]
	
        2023 May 02	&	IXPE	& 02008201		&	0.63	& 20.9 & soft\\	[0.1cm]
        2023 May 09	&   IXPE	& 02008301		&	 0.87	& 31.0 & soft\\	[0.1cm]
        2023 May 24	&	IXPE	& 02008401		&	0.55	& 24.8 & soft\\	[0.1cm]
 					& 	NICER	& 6643010101	&				& 4.7  & soft\\	[0.04cm]
 					& 	  		& 6643010102	&				& 2.8  & soft\\	[0.04cm]
					&   NuSTAR  & 80902318002	&				& 13.8 & soft\\ 	[0.1cm]
        2023 June 13	&	IXPE	& 02008501		&	0.13	& 28.7 & soft\\	[0.04cm]
					&   NuSTAR	& 80902318003	&				& 0.2  & soft\\	[0.04cm]
					&   	& 80902318004	&				& 9.3  & soft\\	[0.1cm]
        2023 June 20		&	IXPE	& 02008601		&	 0.38	& 34.6 & soft\\	[0.04cm]
					& 	NICER	& 6643010103	&				& 20.1  & soft\\	[0.04cm]
					& 			& 6643010104	&				& 7.6  & soft\\	[0.04cm]
					&   NuSTAR  & 80902318006 	&				& 10.5 & soft \\	[0.04cm]
        \hline \\   
    \end{tabular}
\end{table}


\section{Results}

\subsection{Spectral analysis}
\begin{figure}[h!]
    \centering
    \begin{minipage}[t]{0.5\textwidth}  
        \raggedright 
        \includegraphics[width=\textwidth]{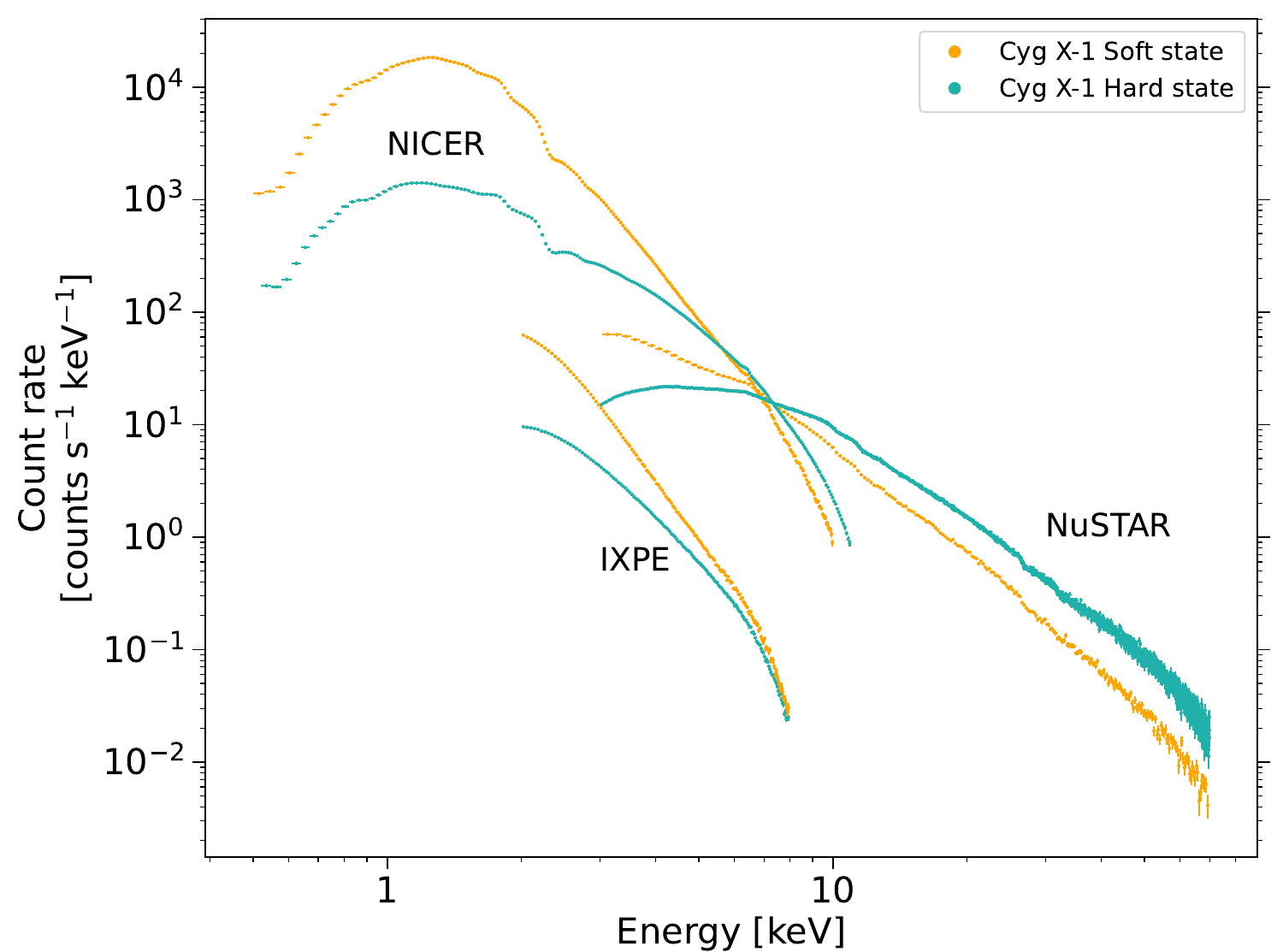}
    \end{minipage}
    \hspace{0.001\textwidth}  
    \begin{minipage}[t]{0.48\textwidth} 
        \centering 
        \includegraphics[width=1.06\textwidth]{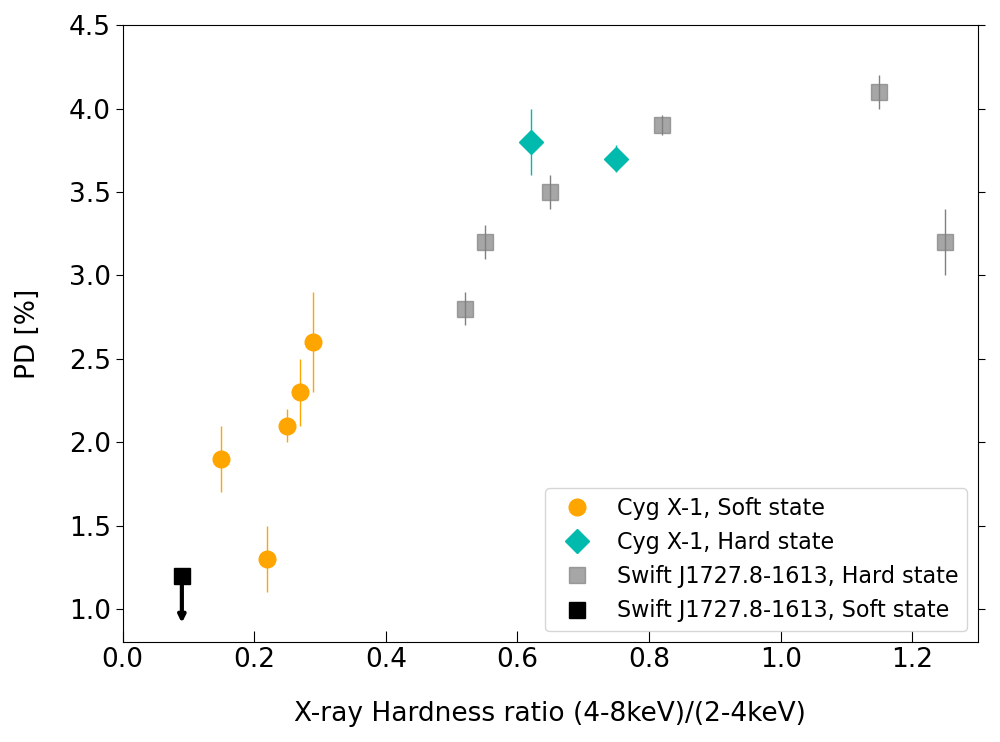}
    \end{minipage}
        \caption{Left: Observed X-ray spectra of NICER, NuSTAR and IXPE data in the soft-intermediate state (orange) and in the hard state (blue) of Cyg X--1. Right: The PD with respect to the hardness ratio defined as the energy flux ratio in the energy bands (4--8 keV)/(2--4 keV) obtained from IXPE spectra. The data in the soft and hard states are shown respectively in orange and blue for Cyg X--1 vs in black and grey for Swift J1727.8--1613.}
        \label{fig:pl_lda_PD_HR}
\end{figure}

\indent Figure \ref{fig:pl_lda_PD_HR} shows the observed spectra of Cyg X--1 over the 0.5--70 keV energy range. The spectra from the three missions are shown in different colours for the different states. Over the entire energy range (0.5--70 keV), the count rate is maximum at 1.5 keV and minimum at 70 keV. Below 8 keV, the count rate is on average higher in the soft state than in the hard state. Above 8 keV, it is the opposite. Moreover, the spectra can be modelled by a thermal blackbody below 8 keV and a power law above 8 keV.

Swift J1727.8--1613 shows similar spectral properties to Cyg X--1, with a higher count rate in the soft state \cite[see][]{2024ApJ...966L..35S} than in the hard state \cite[see][]{2023ApJ...958L..16V, 2024ApJ...960L..17P, 2024ApJ...968...76I} for the same three missions. In the hard state, the spectra show a hard X-ray tail with a high-energy cutoff above 10 keV and a reflection feature at 6.4 keV. Moreover, Swift J1727.8--1613’s spectra show an additional high-energy cutoff at even higher energy than the hard X-ray tail, which is not related to the reflection component \cite[][]{2024ApJ...960L..17P}.

\subsection{Polarization analysis of Cyg X--1 and Swift J1727.8--1613}
Swift J1727.8--1613 shows similar polarization properties to Cyg X--1. The right side of Figure \ref{fig:pl_lda_PD_HR} shows the PD in the soft-intermediate state and in the hard state as a function of the hardness ratio (HR). The PD is calculated from the Stokes parameters Q and U in IXPEOBSSIM over the entire energy range 2--8 keV covered by IXPE. The spectral Stokes parameters I are then used to estimate the HR defined as the ratio between the energy flux in the energy ranges 4--8 keV over 2--4 keV. For each IXPE observation, we summed the PDs of the 3 detector units.
In the soft state, Cyg X--1 has a higher HR and PD than Swift J1727.8--1613. In this state, Cyg X--1's PD reaches a maximum value of 2.6\% for an HR of approximately 0.26. Swift J1727.8--1613 has an upper limit on the PD of 1.3\% for a similar HR of roughly 0.10. In the hard state, Swift J1727.8–1613’s PD varies between 2.7\% and 4\% and the HR varies between 0.5 and 1.25. Cyg X--1's PD stays near 3.7\% for an HR of approximately 0.72 and 0.62 for the first and the second observations, respectively. For both sources, the PD increases with the HR in both spectral states, with significantly higher PDs in the hard state than in the soft state. 

On the other hand, the optical polarization in the V band shows an intrinsic PD of approximately 0.82\% for Cyg X--1 and lower than 0.5\% for Swift J1727.8--1613 in the hard state \cite[see][]{2023ATel16245....1K}.

\section{Discussion and conclusions}

In the hard state, the emission is dominated by the Comptonized emission from the hot plasma. Thus, part of the accretion power is dissipated in the highly Comptonized medium \cite[][]{2007A&ARv..15....1D}. On the other hand, the soft state is dominated by the thermal emission from the standard accretion disk. Nevertheless, the Comptonized emission from the corona can still contribute to 10-35\% of the total emission of the system in Cyg X--1 \cite[][]{1999MNRAS.309..496G, 2014ApJ...780...78T}. Thus, Cyg X--1 still contains part of the Comptonized emission from the corona. For this reason, in the soft state, Cyg X--1 has a harder spectrum and a higher PD than Swift J1727.8--1613. Therefore, despite Cyg X--1 being a HMXB, in the 2--8 keV energy range, we conclude that it exhibits similar X-ray polarization properties to the LMXB Swift J1727.8--1613, with a similar evolution of the PD with the X-ray spectral hardness. \\
According to \cite{1960ratr.book.....C} and \cite{1963trt..book.....S}, the polarization from
the Comptonized emission comes from the multiple scattering on the relativistic electrons. Therefore, the PA should be perpendicular to the scattering plane, depending on the system’s orientation and the corona geometry \cite[][]{2010ApJ...712..908S}. Accordingly, in the hard state, the PAs from the Comptonized and thermal emission should be orthogonal to each other, which depolarizes the system.
Nevertheless, for highly spinning black holes in the soft state, part of the radiation emitted from the disk will return to the accretion disk because of the strong gravitational effect (which produces light bending), and part of it will be reflected off the disk surface \cite[][]{2009ApJ...701.1175S}. As a consequence, the PD is high \cite[][]{2010ApJ...712..908S, 2024ApJ...969L..30S} and the PA is directed along the black hole rotational axis, which is also the radio jet's axis \cite[][]{2024ApJ...971L...9W, 2024ApJ...966L..35S, 2022Sci...378..650K}. These results match the IXPE observations and modelling performed by \cite{2024ApJ...966L..35S} in the soft state of Swift J1727.8--1613 (see their Figure 6). They modelled the PD and PA using a relativistic accretion disk model for different spins, inclinations, and proportions of the incoming emission from the disk reflected on the disk's surface (also called the albedo). For the system to match the IXPE observations -- high PD and a PA aligned with the radio jet -- the corona needs to sandwich the standard accretion disk \cite[i.e. slab corona geometry, see][]{2022Sci...378..650K, 2023ApJ...949L..10P, 2024ApJ...969L..30S}. \\
\noindent In conclusion, the observed results from Cyg X--1 and Swift J1727.8--1613 support an X-ray polarization produced in the innermost regions of the accretion disk, with a radially extended corona. The X-ray polarization is intrinsically linked to the Comptonized emission in the hard state and to the returning radiation from the disk in the soft state.\\
\indent On the other hand, the decrease in the intrinsic optical PD shows that less radiation is scattered, which might indicate the presence of a weaker stellar outflowing material in the LMXB Swift J1727.8--1613 than for the HMXB Cyg X--1 \cite[][]{2023A&A...678A..58K, 2024A&A...682L...1M}. The optical polarization differences between the two systems emphasize the different nature of the donors. However, we observed similarities in the X-ray polarization properties. Therefore, the discrepancy between the optical and X-ray polarization indicates that the innermost accretion processes are independent of the companion star's type or the presence of a strong stellar wind. \\

\textbf{Acknowledgments:} 
We thank J. Steiner for providing the reduced NICER data and V. Kravtsov for valuable discussion. M.B. acknowledges the support from GAUK project No. 102323. J.S. thanks GACR project 21-06825X for the support.

%
\hypersetup{hidelinks}
\bibliographystyle{mnras}
\input{main.bbl}

\end{document}